# A Mechanism for Detection of Gray Hole Attack in Mobile Ad Hoc Networks

Jaydip Sen, M. Girish Chandra, Harihara S.G., Harish Reddy, P. Balamuralidhar
Embedded Systems Research Group, Tata Consultancy Services,
#96, EPIP Industrial Estate, Whitefield Road
Bangalore-560066, INDIA.
{jaydip.sen, m.gchandra, harihara.g, h.reddy, balamurali.p}@tcs.com

*Abstract*—Protecting the network layer from malicious attacks is an important and challenging security issue in mobile ad hoc networks (MANETs). In this paper, a security mechanism is proposed to defend against a cooperative gray hole attack on the well known AODV routing protocol in MANETs. A gray hole is a node that selectively drops and forwards data packets after it advertises itself as having the shortest path to the destination node in response to a route request message from a source node. The proposed mechanism does not apply any cryptographic primitives on the routing messages. Instead, it protects the network by detecting and reacting to malicious activities of any node. Simulation results show that the scheme has a significantly high detection rate with moderate network traffic overhead.

## I. INTRODUCTION

A Mobile Ad Hoc Network (MANET) is a group of mobile nodes that cooperate and forward packets for each other. Such networks extend the limited wireless transmission range of each node by multi-hop packet forwarding, and thus they are ideally suited for scenarios in which pre-deployed infrastructure support is not available. MANETs have some special characteristic features such as unreliable wireless links used for communication between hosts, constantly changing network topologies, limited bandwidth, battery power, low computation power etc. While these characteristics are essential for the flexibility of MANETs, they introduce specific security concerns that are either absent or less severe in wired networks. MANETs are vulnerable to various types of attacks including passive eavesdropping, active interfering, impersonation, and denial-of-service. Intrusion prevention measures such as strong authentication and redundant transmission should be complemented by detection techniques to monitor security status of these networks and identify malicious behavior of any participating node(s). One of the most critical problems in MANETs is the security vulnerabilities of the routing protocols. A set of nodes may be compromised in such a way that it may not be possible to detect their malicious behavior easily. Such nodes can generate new routing messages to advertise non-existent links, provide incorrect link state information, and flood other nodes with routing traffic, thus inflicting Byzantine failure in the network. In this paper, we discuss one such attack known as *Gray Hole Attack* on the widely used AODV (Ad hoc On-demand Distance Vector) routing protocol in MANETs. A mechanism is presented to detect and defend the network against such an attack which may be launched cooperatively by a set of malicious nodes.

The rest of the paper is organized as follows. Section II discusses some related work on routing security in MANETs. Section III defines and discusses various types of gray holes attacks on MANETs. Section IV describes the details of the proposed mechanism for detection of gray hole nodes. Section V presents the simulation conducted on the proposed mechanism and the performance analysis of the scheme. Section VI concludes the paper while highlighting some future scope of work.

## II. RELATED WORK

The problem of security and cooperation enforcement has received considerable attention by researchers in the ad hoc network community. In this section, we discuss some of these works. Mechanisms for securing the routing layer of a MANET by cryptographic techniques are proposed by Hu et al [1], Papadimitratos and Hass [2], Sanzgiri et al [3] and Yang et al [10]. Schemes to handle authentication in ad hoc networks by trusted certificates authorities (CAs) have been proposed by Zhou and Haas [4]. Hubaux et al [5] have proposed a self-organized PGP-based scheme to authenticate nodes using chains of certificates and transitivity of trust. Some researchers have also focused on detecting and reporting misleading routing misbehavior of nodes. *Watchdog* and *Pathrater* [6] use observation-based techniques to detect misbehaving nodes and report observed misbehavior back to the source of the traffic. However, the scheme does not punish malicious nodes; instead, they are relieved of their packet forwarding burden. Researchers have also investigated means of discouraging selfish routing behavior in ad hoc networks, generally through payment schemes [7]. These approaches either require the use of tamper-proof hardware or central bankers to do the accounting securely, both of which may not be appropriate in some truly ad hoc network scenarios. Deng, Li and Agrawal [8] have suggested a mechanism of defense against a black hole attack on AODV routing protocol. In their proposed scheme, when the *RouteReply* packet is received from one of the intermediate nodes, another *RouteRequest* is sent from the source node to the neighbor node of the intermediate node in the path. This is to check whether such a path really exists from the intermediate node to the destination node. While this scheme completely eliminates the black hole attack by a single attacker, it fails miserably in identifying a cooperative black hole attack involving multiple malicious nodes.

## III. GRAY HOLE ATTACK

We first present a security vulnerability in the AODV protocol, and then describe different types of gray hole attacks. In AODV protocol, every mobile node maintains a routing table that stores the next hop node information for a route to a destination node. When a source node wishes to route a packet to a destination node, it uses the specified route if such a route is available in its routing table. Otherwise, the node initiates a route discovery process by broadcasting a *RouteRequest* (RREQ) message to its neighbors. On receiving a RREQ message, the intermediate nodes update their routing tables for a reverse route to the source node. All the receiving nodes that do not have a route to the destination node broadcast the RREQ packet to their neighbors. Intermediate nodes increment the hop count before forwarding the RREQ. A *RouteReply* (RREP) message is sent back to the source node when the RREQ query reaches either the destination node itself or any other node that has a current route to the destination.

We now describe the gray hole attack on MANETS. The gray hole attack has two phases. In the first phase, a malicious node exploits the AODV protocol to advertise itself as having a valid route to a destination node, with the intention of intercepting packets, even though the route is spurious. In the second phase, the node drops the intercepted packets with a certain probability. This attack is more difficult to detect than the black hole attack where the malicious node drops the received data packets with certainty [8]. A gray hole may exhibit its malicious behavior in different ways. It may drop packets coming from (or destined to) certain specific node(s) in the network while forwarding all the packets for other nodes. Another type of gray hole node may behave maliciously for some time duration by dropping packets but may switch to normal behavior later. A gray hole may also exhibit a behavior which is a combination of the above two, thereby making its detection even more difficult.

## IV. PROPOSED MECHANISM

In this section, we first mention some practical assumptions that have been made for formulating the network model and then present the proposed mechanism in detail.

### A. Network Model

We consider a MANET consisting of similar types of nodes. Each node may freely roam, or remain stationary in a location for an arbitrary period of time. In addition, each node may join or leave the network, or fail at any time. The nodes perform peer-to-peer communication over shared, bandwidth-constrained, error-prone, and multi-hop wireless channel. For the purpose of differentiation, we assume that each node has a unique nonzero ID. All the links in the network are assumed to be bi-directional. However, unlike most of the current security frameworks for MANETs, the proposed mechanism does not assume promiscuous mode of operation of the wireless interfaces of the nodes. The promiscuous mode may not only incur extra computation overhead and energy consumption in order to process the transit packets, but also it will not be feasible in cases where the nodes are equipped with directional antennas. There may be varying number of gray hole nodes in the network at different points of time and these malicious nodes may cooperate with each other to disrupt the communication in the network.

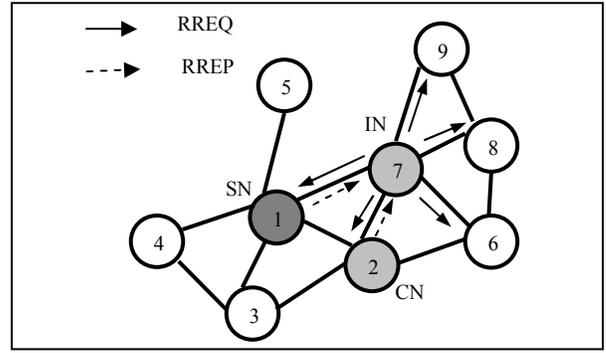

Figure 1. Topology of a MANET

TABLE I. DRI TABLE OF NODE 7

| Node | From | Through | RTS/CTS | CheckBit |
|------|------|---------|---------|----------|
| 1 | 0 | 0 | 15 | 0 |
| 2 | 1 | 1 | 5 | 1 |
| 6 | 0 | 1 | 3 | 0 |
| 8 | 1 | 0 | 6 | 1 |
| 9 | 0 | 1 | 4 | 0 |

### B. Different Modules of the Scheme

The proposed mechanism involves both local and cooperative detection to identify any malicious gray hole node in the network. Once a node is detected to be really malicious, the scheme has a notification mechanism for sending messages to all the nodes that are not yet suspected to be malicious, so that the malicious node can be isolated and not allowed to use any network resources. The mechanism consists of four security procedures which are invoked sequentially. The security procedures are: (1) Neighborhood data collection, (2) Local anomaly detection, (3) Cooperative anomaly detection, and (4) Global alarm raiser. Each of these procedures is described in details below.

(1) *Neighborhood data collection module*: Each node in the network collects the data forwarding information in its neighborhood and stores it in a table known as the *Data Routing Information* (DRI) table. The DRI table of node 7 in Fig. 1 is shown in Table I. In its DRI table node 7 maintains packet routing information of its neighbor nodes 1, 2, 6, 8, and 9. An entry '1' for a node under the column '*From*' implies that node 7 has forwarded data packet coming from that node and an entry '1' for a node under the column '*Through*' implies that node 7 has forwarded data packets to that node. Thus, as per Table 1, node 7 has neither forwarded any data packet from node 1 nor it has forwarded any data packet to node 1. However, node 7 has forwarded data packets to node 2 and also has forwarded data packets that have come from node 2. In this way, each node constructs its DRI table and maintains it. After a certain threshold time interval, (this depends on the mobility of the network) each node identifies its neighbors with which it has not interacted for the purpose of data communication, and invokes subsequent detection procedures to probe them further. This identification is done on the basis of the nodes that have '0' entries both in the '*From*' and '*Through*' columns in the DRI table. For example, as shown in Table I, node 7 has not communicated to node 1. Therefore, the node 7 invokes the local anomaly detection procedure for node 1. The 'RTS/CTS' column in the DRI table gives the ratio of the number of *request to send* (RTS) messages to the number of *clear to send* (CTS) messages for the corresponding node. This gives a rough idea about the number of requests arriving at the node for data communication and number packet transmission that the node is actually doing. The significance of the column '*CheckBit*' in the DRI table will be discussed in local anomaly detection procedure.

TABLE II. PROBE CHECK TABLE FOR NODE 7

| NodeID | ProbeStatus |
|--------|-------------|
| 2 | 0 |
| 6 | 1 |
| 8 | 1 |
| 9 | 1 |

(2) *Local anomaly detection module*: This security procedure is invoked by a node when it identifies a suspicious node by examining its DRI table as discussed above.

We call the node that initiates the local anomaly detection procedure as the *Initiator Node* (IN). The IN first chooses a *Cooperative Node* (CN) in its neighborhood based on its DRI records and broadcasts a RREQ message to its 1-hop neighbors requesting for a route to the CN. In reply to this RREQ message the IN will receive a number of RREP messages from its neighboring nodes. It will certainly receive a RREP message from the *Suspected Node* (SN) if the latter is really a gray hole (since the gray holes always send RREP messages but drop data packets probabilistically). After receiving the RREP from the SN, the IN sends a probe packet to the CN through the SN. After the *time to live* (TTL) value of the probe packet is over, the IN enquires the CN whether it has received the probe packet. If the reply to this query is affirmative, (i.e., the probe packet is really received by the CN) then the IN updates its DRI table by making an entry '1' under the column '*CheckBit*' against the node ID of the SN. However, if the probe packet is found to have not reached the CN, the IN increases its level of suspicion about the SN and activates the cooperative anomaly detection procedure, as discussed later in this Section.

In Fig. 1, node 7 acts as the IN and initiates the local anomaly detection procedure for the SN (node 1) and chooses node 2 as the CN. Node 2 is the most reliable node for node 7 as both the entries under columns 'From' and 'Through' for node 2 are '1'. Node 7 broadcasts a RREQ message to all its neighbor nodes 1, 2, 6, 8 and 9 requesting them for a route to the CN, i.e., node 2 in the example. After receiving a RREP from the SN (node 1), node 7 sends a probe packet to node 2 via node 1. Node 7 then enquires node 2 whether it has received the probe packet. If node 2 has received the probe packet, node 7 makes an entry '1' under the column '*CheckBit*' in its DRI table corresponding to the row of node 1. If node 2 has not received the probe packet, then node 7 invokes the cooperative anomaly detection procedure.

(3) *Cooperative anomaly detection module*: The objective of this procedure is to increase the detection reliability by reducing the probability of false detection of local anomaly detection procedure.

This procedure is activated when an IN observes that the probe packet it had sent to the CN through the SN did not reach the CN. The IN invokes the cooperative detection procedure and sends a cooperative detection request message to all the neighbors of the SN. When the neighbors of the SN receive the cooperative detection request message, each of them sends a RREQ message to the SN requesting for a route to the IN. After the SN responds with a RREP message, each of the requesting nodes sends a 'further probe packet' to the IN along that route. This route will obviously include SN, as SN is a neighbor of each requesting node and the IN as well. Each neighbor of the SN (except the IN) now notifies the IN that a 'further probe packet' has already been sent to it. This 'notification message' from each neighbor is sent to the IN through routes which do not include the SN. This is necessary to ensure that the SN is not aware about the on-going cross checking process. The IN will receive numerous 'further probe packets' and 'notification messages'. The IN now constructs a *ProbeCheck* table. The ProbeCheck table has two fields: *NodeID* and *ProbeStatus*. Under the NodeID field, the IN enters the identifiers of the nodes which have sent notification messages to it. An entry of '1' is made under the column 'ProbeStatus' corresponding to the nodes from which the IN has received the 'further probe packet'.

An example ProbeCheck table for node 7 of the network in Fig.1 is presented in Table II. It can be observed that node 7 has received the 'further probe packet' from all the neighbors of the SN (node 1) except node 2. There may be a possibility that the probe packet might have not been maliciously dropped by the SN, rather it has been lost because of collision or buffer overflow. A mathematical estimation can be made for the probability of collision or buffer overflow at the SN [9]. However, to avoid complex mathematical computation, we propose a simple mechanism where each node sends three 'further probe packets' interspaced with a small time interval. If none of these three packets from a neighbor are received by the IN, the SN is believed to be behaving like a gray hole for that node during that time. This gray hole behavior may be exhibited for a single node (as for node 2 in Table II) or may be for a group of nodes.

If the SN is found to behave like a gray hole, it is isolated from the network by invocation of the global alarm detection procedure as discussed later in this section. The frequency of invocation of the detection algorithm is important for ensuring the desired throughput in the network as a gray hole may quickly change its phase form 'good' to 'bad'. The periodicity of invocation of the algorithm should be based on the maximum percentage of packet drop that the network application can afford. In the worst case, a gray hole will just change its phase from 'good' to 'bad' immediately after the invocation of one round of the detection algorithm is over and will switch back to 'good' phase just before the next invocation. Although such a situation may be quite unlikely, the invocation frequency should be based on the estimation of the number of packets that the gray hole may drop during that period and the maximum number of packet drops that may be allowed while still maintaining the desired quality of service (QoS).

(4) *Global alarm raising module*: This procedure is invoked to establish a network wide notification system for sending alarm messages to all the nodes in the network about the gray hole node(s) that has been detected by the cooperative anomaly detection algorithm. It also ensures that the identified malicious node(s) is isolated so that it cannot use any network resources.

The detection and isolation mechanism of malicious nodes may involve a security problem. A group of malicious nodes can collude together to launch a *bad mouthing attack* by falsely accusing a legitimate node and isolating it from the network. To prevent this, we propose a mechanism that is similar to threshold cryptography [4]. In the proposed mechanism, when a node identifies a suspected node to be malicious by invocation of the cooperative detection procedure, it sends a digitally signed (using its private key) alarm message to all its neighbors. The full signature is constructed when at least $k$ nodes put their signatures into the alarm message. Once the alarm message is authenticated with the full signature, the suspected node is isolated from the network. Thus the proposed scheme is robust against collusion involving maximum $k$-1 malicious nodes in a neighborhood. When a node is finally identified to be malicious, its node ID will be entered into a global list of malicious nodes called '*faulty list*'. The faulty list is periodically flooded in the network, as and when an update is made into it. In order to reduce the overhead, we propose two alternative methods of propagation of faulty list. The faulty list may be piggybacked with the normal routing messages (RREQ and RREP messages) so that it does not cause any extra overhead. Alternatively, each node may maintain a partial list of faulty nodes which are in its immediate neighborhood. This partial list is updated as and when the neighborhood of the node changes. This mechanism will be particularly suitable for routing protocols like AODV, as nodes need to know the information regarding its neighboring nodes only for routing.

## V. SIMULATION AND RESULTS

The proposed mechanism is implemented in network simulator *ns-2* for the purpose of evaluation of its performance. The physical layer at each networking interface is chosen to approximate the Lucent WaveLAN wireless card. The MAC layer protocol and the routing protocol are 802.11 DCF and AODV respectively. An improved version of 'random waypoint' is used as the mobility mode [11]. The host pause time is chosen to be zero to simulate a continuously mobile network. Malicious gray holes are simulated using a two-phase *Markov Chain Machine*. While in the good phase none of the gray holes drop any packet, in the bad phase, packets are dropped based on a function that generates a random number between a maximum value (MAX_RATE) and a minimum value (MIN_RATE). The simulation parameters are presented in Table III. The metrics that are used for evaluating the proposed mechanism are: (1) False positive rate, i.e., the probability of incorrectly identifying a legitimate node as malicious, (2) Misdetection (false negative) rate, i.e., the probability of failure in detecting a malicious node, (3) Data packet delivery ratio, i.e., the percentage of data packets that are successfully delivered, and (4) Communication overhead due to control packets of the security algorithm.

TABLE III. SIMULATION PARAMETERS

| Parameter | Value |
|---|---|
| Simulation area | 2000m * 600m |
| Simulation duration | 1500 s |
| No. of mobile nodes | 50 |
| Transmission range | 200 m |
| Movement model | Random waypoint |
| Maximum speed of node | 20 m/s |
| Traffic type | CBR (UDP) |
| Total no. of flows | 20 |
| Packet rate | 2 packets/s |
| Data payload | 512 bytes/packet |
| Max no. of malicious nodes | 10 |
| Host pause time | 0 |

Fig. 2 shows how false positive rate varies with mobility of the nodes for different percentages of gray hole nodes in the network. The maximum value of observed false positive rate is 7%. The false positive rate increases as the nodes move faster. If a node constantly moves at a high speed, it can gather only partial information about its transmissions with its current neighbors. As a result, it is more likely to make mistakes.

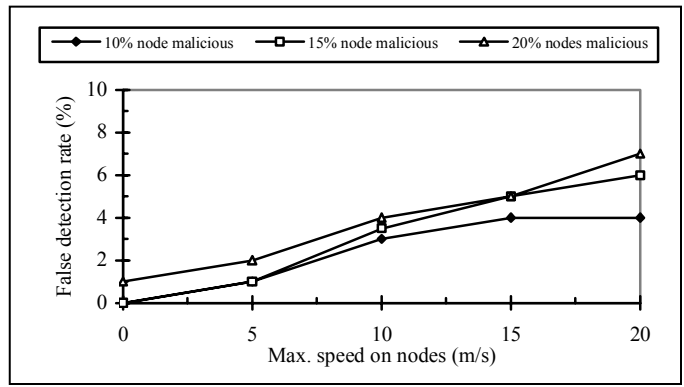

Figure 2. False positive rate vs. nodes' mobility

Fig. 3 depicts the variation of miss-detection rate with nodes' mobility for different percentages of gray hole nodes. It is seen that misdetection ratio is maximum (12%) in a static network and starts dropping with the mobility of the nodes. This is because in a static network if a gray hole remains in a sparsely populated region, its neighbors may not be able to punish it since there may not be requisite $k$ number of nodes to arrive at the consensus. On the contrary, in a mobile network, the mobility increases the probability that other nodes roam into the region of the gray hole node(s), or the gray hole node(s) enters into a densely populated region. As a result, it is less likely that a gray hole would be able to escape without being detected.

Fig. 4 shows how the data packet delivery ratio varies with respect to the number of gray hole nodes for the normal AODV protocol and for the proposed secure routing protocol. It is observed that even when 20% of the nodes in the network are malicious gray holes, the percentage of packets successfully delivered is more than 90% if the proposed security protocol is applied. However, 100% packet delivery ratio is not achieved even with the proposed security mechanism. A careful analysis of the trace files has shown that most of the packet loss occurs during the detection and reaction phase of the algorithm.

Fig. 5 shows the communication overhead of the control packets of the security mechanism. We have found that with the increase in number of gray holes, the overhead increases. Therefore, we have reported the result for the worst case when the percentage of malicious node in the network is 20%. The communication overhead is shown as the percentage of the number of control packet required to route different number of data packets in the network. The normal AODV performance is taken as the base line. It is observed that the overhead of the proposed mechanism drops as the number of data packets transmitted is increased. This clearly demonstrates that the scheme is efficient in terms of communication overhead.

## VI. CONCLUSION

In this paper, we have presented a mechanism for detection of malicious gray hole nodes in MANETs. Due to their occasional misbehavior, the gray holes are very difficult to detect. The proposed security mechanism increases the reliability of detection by proactively invoking a collaborative and distributed algorithm involving the neighbor nodes of a malicious gray hole node. Detection decision works on a consensus algorithm based on threshold cryptography. The simulation results show that the mechanism is effective and efficient with high detection rate and very low false positive rate and control overhead.

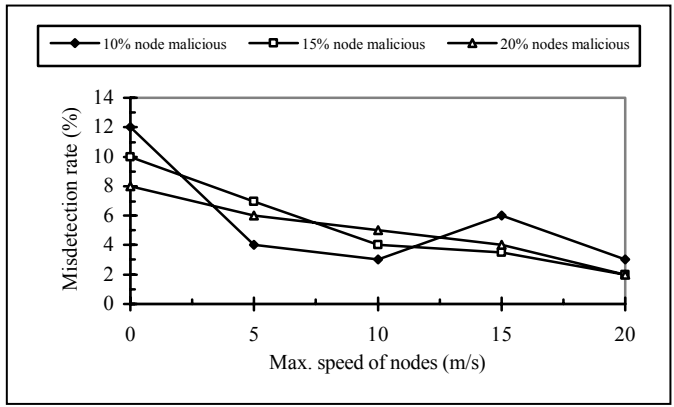

Figure 3. Misdetection rate vs. nodes' mobility

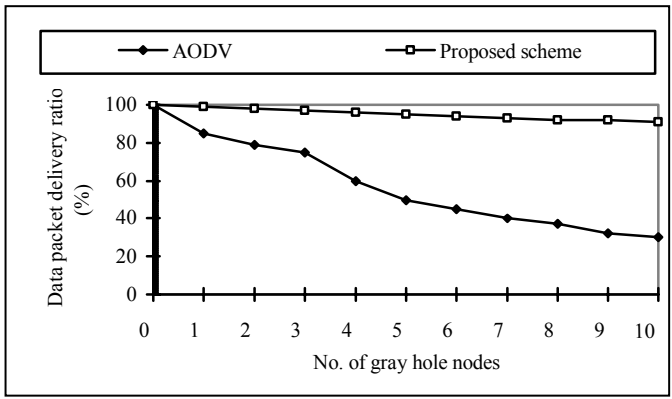

Figure 4. Data packet delivery rate vs. No. of gray hole nodes

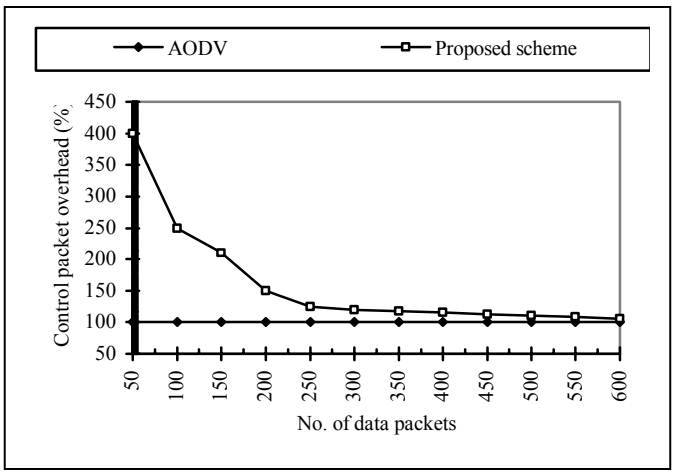

Figure 5. Control packet overhead for different no. of data packets